\documentclass[twocolumn,showpacs,preprintnumbers,amsmath,amssymb]{revtex4}
\usepackage{epsfig}
\usepackage{dcolumn}

\usepackage{graphicx}
\usepackage{dcolumn}
\usepackage{bm}

\newcommand{\be}{\begin{equation}}
\newcommand{\ee}{\end{equation}}
\newcommand{\bea}{\begin{eqnarray}}
\newcommand{\eea}{\end{eqnarray}}

\newcommand{\integer}{\relax{\rm I\kern-.18em N}}

\begin{document}

\title{Moyal dynamics of constraint systems\footnote{
Presented at the XIII Annual Seminar {\it Nonlinear Phenomena in Complex Systems: 
Chaos, Fractals, Phase Transitions, Self-organization}
at Minsk, Belarus, May 16-19, 2006. 
}
}

\author{M. I. Krivoruchenko}

\affiliation{
Institute for Theoretical and Experimental Physics$\mathrm{,}$ 
B. Cheremushkinskaya 25 \\
117259 Moscow, Russia
}

\begin{abstract}
Quantization of constraint systems within the Weyl-Wigner-Groenewold-Moyal 
framework is discussed. Constraint dynamics of classical and quantum systems is 
reformulated using the skew-gradient projection formalism. The quantum deformation 
of the Dirac bracket is generalized to match smoothly the classical Dirac bracket 
in and outside of the constraint submanifold in the limit $\hbar \to 0$.
\end{abstract}


\maketitle

\section{Introduction} 
\setcounter{equation}{0}

Gauge symmetries provide mathematical basis for known fundamental interactions. 
Within the generalized Hamiltonian framework \cite{DIRAC}, 
gauge theories correspond to first-class constraints systems. Upon gauge fixing, 
these systems convert to second-class constraint systems. The operator quantization 
schemes for constraints systems have been developed by Dirac \cite{DIRAC}. 
The path integral quantization has also been developed and 
found to be especially effective for gauge theories (for reviews see \cite{HENN,FASL}).

Besides conventional operator formulation of quantum mechanics and the
path integral method, the popular approach to quantization of
classical systems is based on the Groenewold star-product formalism \cite
{GROE}. It takes the origin from the Weyl's association rule \cite{WEYL} 
between operators in the Hilbert space and functions in phase
space and the Wigner function \cite{WIGNER}. The star-product formalism is 
known also under the names of the deformation quantization and the Moyal 
quantization \cite{MOYAL,BARLE}.

The skew-symmetric part of the star-product, named the Moyal bracket,
governs the evolution of quantum systems in phase space, just like the Poisson
bracket governs the evolution of classical unconstrained systems and the Dirac
bracket governs the evolution of classical constraint systems. The Moyal bracket
represents the quantum deformation of the Poisson bracket. The quantum deformation 
of the Dirac bracket has been constructed recently \cite{KRFA}.

The outline of the paper is as follows: In the next Sect., we give a pedagogical 
introduction to the Weyl's association rule using the elegant method developed 
by Stratonovich \cite{strat} and give an introduction to the star-product formalism.
More details on this subject can be found in articles \cite{BERRY,BAYEN,CARRU,HILL,BALAZ,
VOROS}.

The phase-space functions and the Dirac bracket do not make any physical sense 
outside of constraint submanifolds. In Ref. \cite{KRFA} we constructed the quantum 
deformation of the Dirac bracket on the constraint submanifold, sufficient for the
purpose of generating time evolution of quantum constraint systems. It would, 
however, be interesting from the abstract point of view to have a quantum-mechanical 
extension of the Dirac bracket which matches smoothly at $\hbar \to 0$ with 
the classical Dirac bracket outside of the constraint submanifold also. 

This problem is addressed and solved in Sects. III and IV. In Sect. III, 
we reformulate the classical constraint dynamics using projection formalism 
and present the classical Dirac bracket of functions in terms of the Poisson bracket 
of functions projected onto constraint submanifold. Sect. IV gives the quantum-mechanical
generalization of the method proposed. Sects. III-D and IV-B,C contain new results,
the others is a pedagogical exposition of earlier works (mainly \cite{KRFA}).  

In Conclusion, we summarize results.


\section{Weyl's association rule and the star-product}
\setcounter{equation}{0} 


Systems with $n$ degrees of freedom are described by $2n$ canonical
coordinates and momenta $\xi ^{i}=(q^{1},...,q^{n},p_{1},...,p_{n})$. These
variables parameterize phase space $T_{*}\mathbb{R}^{n}$ defined as the
cotangent bundle of $n$-dimensional configuration space $\mathbb{R}^{n}$.
Canonical variables satisfy the Poisson bracket relations 
\begin{equation}
\{\xi ^{k},\xi ^{l}\}=-I^{kl}.  \label{POIS}
\end{equation}
The skew-symmetric matrix $I^{kl}$ has the form 
\begin{equation}
\left\| I\right\| =\left\| 
\begin{array}{ll}
0 & -E_{n} \\ 
E_{n} & 0
\end{array}
\right\|  \label{MATR}
\end{equation}
where $E_{n}$ is the $n\times n$ identity matrix and imparts to 
$T_{*}\mathbb{R}^{n}$ a skew-symmetric bilinear form. The phase space 
acquires thereby structure of symplectic space. The distance between 
two points in phase space is not defined. One can measure, however, areas 
stretched on any two vectors $\xi^{k} $ and $ \zeta^{l}$ as
$\mathcal{A}=I_{kl} \xi^{k} \zeta^{l}$
where $I_{kl}=-I^{kl}$ so that $I_{kl}I^{lm}=\delta _{k}^{m}$.

Principal similarities and distinctions between Euclidean and symplectic
spaces are cataloguized in Table 1. For skew-gradients of functions,
short notation $Idf(\xi)$ is used.

\begin{table}[tbp]
\caption{Comparison of properties of Euclidean and symplectic spaces }
\label{lab1}$
\begin{array}{|c|c|}
\hline\hline
\mathbf{Euclidean\;space} & \mathbf{Symplectic\;space} \\ 
 x,y \in \mathbb{R}^{n} & \xi, \zeta \in \mathbb{R}^{2n}  \\ \hline\hline
\begin{array}{c}
\mathrm{Metric\;structure} \\ 
g_{ij}=g_{ji}\; \\ 
g_{ij}g^{jk}=\delta _{i}^{k}
\end{array}
& 
\begin{array}{c}
\mathrm{Symplectic\;structure} \\ 
I_{ij}=-I_{ji} \\ 
I_{ij}I^{jk}=\delta _{i}^{k}
\end{array}
\\ \hline
\begin{array}{c}
\mathrm{Scalar\;product\;} \\ 
(x,y)=g_{ij}x^{i}y^{j}
\end{array}
& 
\begin{array}{c}
\mathrm{Skew-scalar\;product\;} \\ 
(\xi ,\zeta )=I_{ij}\xi ^{i} \zeta ^{j}
\end{array}
\\ \hline
\begin{array}{c}
\mathrm{Distance} \\ 
L = \sqrt{(x - y,x - y)}
\end{array}
& 
\begin{array}{c}
\mathrm{Area} \\ 
\mathcal{A} = (\xi ,\zeta )
\end{array}
\\ \hline
\begin{array}{c}
\mathrm{Gradient\;} \\ q
(\bigtriangledown f)^{i} = g^{ij}\partial f/\partial x^{j} \\
\;\\
\end{array}
& 
\begin{array}{c}
\mathrm{Skew-gradient\;} \\ 
(Idf)^{i} \equiv - I^{ij}\partial f/\partial \xi ^{j} \\ 
\;\; = \{ \xi^{i},f \}
\end{array}
\\ \hline
\begin{array}{c}
\mathrm{Scalar\;product} \\ 
\mathrm{of\;gradients\;of} f \mathrm{\;and\;} g \\ 
(\bigtriangledown f, \bigtriangledown g)
\end{array}
& 
\begin{array}{c}
\mathrm{Poisson\;bracket} \\
\mathrm{\;of\;} f \mathrm{\;and\;} g \\ 
(Idf,Idg) = \{f,g\}
\end{array}
\\ \hline
\begin{array}{c}
\mathrm{Orthogonality} \\ 
g_{ij}x^{i}y^{j}=0
\end{array}
& 
\begin{array}{c}
\mathrm{Skew-orthogonality} \\ 
I_{ij} \xi ^{i} \zeta ^{j}=0
\end{array}
\\ \hline\hline
\end{array}
$%
\end{table}

In quantum mechanics, canonical variables $\xi ^{i}$ are associated to
operators of canonical coordinates and momenta $\mathfrak{x}^{i}=(%
\mathfrak{q}^{1},...,\mathfrak{q}^{n},\mathfrak{p}_{1},...,\mathfrak{p}_{n})$
acting in the Hilbert space, which obey the commutation relations 
\begin{equation}
\lbrack \mathfrak{x}^{k},\mathfrak{x}^{l}]=-i\hbar I^{kl}.  \label{COMM}
\end{equation}
The Weyl's association rule extends the correspondence $\xi
^{i}\leftrightarrow \mathfrak{x}^{i}$ to phase-space functions $%
f(\xi )\in C^{\infty}(T_{*}\mathbb{R}^{n})$ and operators $\mathfrak{f}\in
Op(L^{2}(\mathbb{R}^{n}))$. It can be illustrated as follows: 
\begin{eqnarray}
\xi ^{i}\in T_{*}\mathbb{R}^{n} &\longleftrightarrow &\mathfrak{x}^{i}\in
Op(L^{2}(\mathbb{R}^{n}))  \nonumber \\
\{\xi ^{i},\xi ^{j}\} &\longleftrightarrow &-\frac{i}{\hbar }[\mathfrak{x}%
^{i},\mathfrak{x}^{j}]  \nonumber \\
f(\xi )\in C^{\infty }(T_{*}\mathbb{R}^{n}) &\longleftrightarrow &%
\mathfrak{f}\in Op(L^{2}(\mathbb{R}^{n}))  \nonumber
\end{eqnarray}

The set of operators $\mathfrak{f}$ acting in the Hilbert space is closed
under multiplication of operators by $c$-numbers and summation of
operators. Such a set constitutes vector space: 
\[
\left. 
\begin{array}{c}
\left. 
\begin{array}{ccc}
c\times f(\xi ) & \longleftrightarrow & c\frak{f} \\ 
f(\xi )+g(\xi ) & \longleftrightarrow & \frak{f}+\frak{g}
\end{array}
\right\} 
\begin{array}{c}
\mathrm{vector} \\ 
\mathrm{space}
\end{array}
\\ 
\begin{array}{ccc}
\;f(\xi )\star g(\xi ) & \longleftrightarrow & \;\;\frak{fg}
\end{array}
\;\;\;\;\;\;\;\;\;\;\;\;\;\;\;\;\;\;\;
\end{array}
\right\} \mathrm{algebra}\vspace{3mm} 
\]
Elements of basis of such a vector space can be labelled by canonical 
variables $\xi^{i}$. The commonly 
used Weyl's basis looks like
\begin{eqnarray}
\mathfrak{B}(\xi ) &=& (2\pi \hbar )^{n}\delta^{2n}(\xi - \mathfrak{x}) 
\nonumber \\
&=& \int \frac{d^{2n}\eta }{(2\pi \hbar )^{n}} \exp (-\frac{i}{\hbar }\eta
_{k}(\xi - \mathfrak{x})^{k}).  \label{P}
\end{eqnarray}

The objects $\mathfrak{B}(\xi )$ satisfy relations \cite{KRFA} 
\begin{eqnarray}
\frak{B}(\xi )^{+} &=&\frak{B}(\xi ),  \nonumber \\
Tr[\frak{B}(\xi )] &=&1,  \nonumber \\
\int \frac{d^{2n}\xi }{(2\pi \hbar )^{n}}\frak{B}(\xi ) &=&\frak{1}, 
\nonumber \\
\int \frac{d^{2n}\xi }{(2\pi \hbar )^{n}}\frak{B}(\xi )Tr[\frak{B}(\xi )%
\frak{f}] &=&\frak{f},  \nonumber \\
Tr[\frak{B}(\xi )\frak{B}(\xi ^{\prime })] = (2\pi \hbar )^{n}\delta
^{2n}(\xi &-&\xi ^{\prime }),  \nonumber \\
\frak{B}(\xi )\exp (-\frac{i\hbar }{2}\mathcal{P}_{\xi \xi ^{\prime }})\frak{%
B}(\xi ^{\prime }) &=&  \nonumber \\
= (2\pi \hbar )^{n}\delta ^{2n}(\xi - \xi ^{\prime })\frak{B}(\xi ^{\prime
}).&\;&  \nonumber
\end{eqnarray}
Here, 
\[
\mathcal{P}_{\xi \xi ^{\prime }} = -I^{kl} \frac{\overleftarrow{\partial }} {%
\partial \xi ^{k} } \frac{\overrightarrow{\partial }}{\partial \xi^{\prime l
}} 
\]
is the so-called Poisson operator.

The Weyl's association rule for a function $f(\xi )$ and an operator $%
\mathfrak{f}$ has the form \cite{strat}
\begin{eqnarray}
f(\xi ) &=&Tr[\mathfrak{B}(\xi )\mathfrak{f}],  \label{S} \\
\mathfrak{f} &=&\int \frac{d^{2n}\xi }{(2\pi \hbar )^{n}}f(\xi )%
\mathfrak{B}(\xi ).  \label{INV}
\end{eqnarray}
In particular, 
\begin{eqnarray}
\xi ^{i} &=&Tr[\mathfrak{B}(\xi )\mathfrak{x}^{i}]  \label{COORDI1} \\
\mathfrak{x}^{i} &=&\int \frac{d^{2n}\xi }{(2\pi \hbar )^{n}}\xi ^{i}%
\mathfrak{B}(\xi ).  \label{COORDI2}
\end{eqnarray}
The function $f(\xi )$ can be treated as the coordinate of $\mathfrak{f}$ in
the basis $\mathfrak{B}(\xi )$, while the right side of Eq.(\ref{S}) can be 
interpreted as the scalar product of $\mathfrak{B}(\xi )$ and $\mathfrak{f}$.

Alternative operator bases and their relations are discussed in Refs. \cite
{BALAZ,MEHTA}. One can make, in particular, operator transforms on 
$\mathfrak{B}(\xi )$ and $c$-number transforms on $\xi ^{i}$. Ambiguities in 
the choice of operator basis are connected to ambiguities in quantization 
of classical systems, better known as "operator ordering problem".

The set of operators is closed under multiplication of operators. The
vector space of operators is endowed thereby with an associative algebra structure.
Given two functions $f(\xi )=Tr[\mathfrak{B}(\xi )\mathfrak{f}]$ and $g(\xi
)=Tr[\mathfrak{B}(\xi )\mathfrak{g}]$, one can construct a third function 
\begin{equation}
f(\xi )\star g(\xi )=Tr[\mathfrak{B}(\xi )\mathfrak{fg}].  \label{GR}
\end{equation}
This operation is called star-product. It has been introduced by Groenewold 
\cite{GROE}. The explicit form of the star-product is as follows: 
\begin{equation}
f(\xi )\star g(\xi )=f(\xi )\exp (\frac{i\hbar }{2}\mathcal{P})g(\xi ),
\label{EG}
\end{equation}
where $\mathcal{P}=\mathcal{P}_{\xi \xi }$.

The star-product splits into symmetric and skew-symmetric parts 
\begin{equation}
f\star g=f\circ g+\frac{i\hbar }{2}f\wedge g.  \label{STAR}
\end{equation}
The skew-symmetric part $f\wedge g$ is known under the name of Moyal
bracket. It is essentially unique \cite{MEHTA}. It governs quantum evolution
in phase space and endows the set of functions with the Poisson algebra
structure: 
\begin{equation}
\begin{array}{c}
\mathrm{physical\;observables} \\ 
\Updownarrow \\ 
\mathrm{functions \; in \; phase \; space} \\ 
\Updownarrow \\ 
\overbrace{\underbrace{\underbrace{f+g,\;c \times f,}_{\mathrm{vector\;space}%
}\;\;f \star g}_{\mathrm{algebra}}, \; f \wedge g}^{\mathrm{Poisson\;algebra}%
}
\end{array}
\label{SCHEME}
\end{equation}

The average values of a physical observable described by function $f(\xi)$
are calculated in terms of the Wigner function 
\begin{equation}
W(\xi )=Tr[\mathfrak{B}(\xi )\mathfrak{r}].  \label{WIGN}
\end{equation}
It is normalized to unity 
\begin{equation}
\int \frac{d^{2n}\xi }{(2\pi \hbar )^{n}}W(\xi )=1.  \label{WFUN}
\end{equation}
If $\mathfrak{f} \leftrightarrow f(\xi)$ and $\mathfrak{r} \leftrightarrow
W(\xi)$ where $\mathfrak{r}$ is the density matrix, then 
\begin{eqnarray}
Tr[\mathfrak{fr}] &=& \int \frac{d^{2n}\xi }{(2\pi \hbar)^{n}}f(\xi )\star
W(\xi )  \nonumber \\
&=& \int \frac{d^{2n}\xi }{(2\pi \hbar)^{n}}f(\xi )W(\xi ).  \label{TR}
\end{eqnarray}
Under the sign of integral, the star-product can be replaced with the
pointwise product \cite{strat}.

Real functions in phase space stand for physical observables, which
constitute in turn the Poisson algebra. If the associative product $f \star
g $ does not commute, its skew-symmetric part gives automatically the
skew-symmetric product which satisfies the Leibniz' law 
\begin{equation}
f\wedge (g\star h)=(f\wedge g)\star h+g\star (f\wedge h).  \label{LEIB}
\end{equation}
This equation is valid separately for symmetric and skew-symmetric parts of
the star-product. In the last case, Eq.(\ref{LEIB}) provides the Jacobi
identity. The validity of the Leibniz' law allows to link the Moyal bracket
with time derivative of functions and build up thereby an evolution equation
for functions in phase space.

In classical limit, the Moyal bracket turns to the Poisson bracket: 
\[
\lim_{\hbar \to 0}f \wedge g = \{f,g\}. 
\]


\section{Classical constraint systems in phase space}
\setcounter{equation}{0} 


Second-class constraints $\mathcal{G}_{a}(\xi) = 0$ with $a = 1,...,2m$ and $m<n$
have the Poisson bracket relations which form a non-degenerate $2m \times 2m$
matrix 
\begin{equation}
\det\{\mathcal{G}_{a}(\xi),\mathcal{G}_{b}(\xi)\} \ne 0.  \label{NONGEN}
\end{equation}
If this condition is not fulfilled, it would mean that gauge degrees of
freedom appear in the system. After imposing gauge-fixing conditions, we
could arrive at inequality (\ref{NONGEN}). Alternatively, breaking condition 
(\ref{NONGEN}) would mean that constraint functions are dependent. After 
removing redundant constraints, we arrive at inequality (\ref{NONGEN}).

Constraint functions are equivalent if they describe the same constraint
submanifold. Within this class one can make 
transformations without changing dynamics.

\subsection{Symplectic basis for constraint functions}

For arbitrary point $\xi$ of constraint submanifold 
$\Gamma^{*} = \{\xi: \mathcal{G}_{a}(\xi) = 0 \}$, 
there is a neighborhood where one may find equivalent constraint functions in terms
of which the Poisson bracket relations look like 
\begin{equation}
\{\mathcal{G}_{a}(\xi),\mathcal{G}_{b}(\xi)\}=\mathcal{I}_{ab}  \label{SB}
\end{equation}
where 
\begin{equation}
\mathcal{I}_{ab}=\left\| 
\begin{array}{ll}
0 & E_{m} \\ 
-E_{m} & 0
\end{array}
\right\|.  \label{SMAT}
\end{equation}
Here, $E_{m}$ is the identity $m\times m$ matrix, $\mathcal{I}_{ab}\mathcal{I%
}_{bc}=-\delta _{ac}$.

The global existence of symplectic basis (\ref{SB}) is an opened
question in general case. The basis (\ref{SB}) always exists locally, i.e.,
in a finite neighborhood of any point of the constraint submanifold. This is
sufficient for needs of perturbation theory. The formalism presented in
this section can therefore to be used to formulate evolution problem of any 
second-class constraints system in phase space in the sense of the
perturbation theory. 

The existence of the local symplectic basis (\ref{SB}) is on the line with
the Darboux's theorem (see, e.g., \cite{ARNO}) which states that in
symplectic space around any point $\xi$ there exists coordinate system in 
$\Delta_{\xi}$ such that $\xi \in \Delta_{\xi}$ where symplectic structure
takes the standard canonical form. Symplectic spaces can be covered by such
coordinate systems.

This is in contrast to Riemannian geometry where metric
tensor at any given point $x$ can always be made Minkowskian, but in any
neighborhood of $x$ the variance of the Riemannian metric with the
Minkowskian metric is, in general, $\sim \Delta x^2$. Physically, by
passing to inertial coordinate frame one can remove gravitation fields at
any given point, but not in an entire neighborhood of that point. The
Darboux's theorem states, reversely, that the symplectic structure can be
made to take the standard canonical form in an entire neighborhood $%
\Delta_{\xi}$ of any point $\xi$. In Riemannian spaces,
locally means at some given point. In symplectic spaces, locally means at
some given point and in an entire neighborhood of that point.

Locally, all symplectic spaces are indistinguishable. Conditionally, one can say 
that any surface in symplectic space, including any constraint surface, is a plane.

In the view of this marked dissimilarity, the validity of Eqs.(\ref{POIS})
in a finite domain looks indispensable.

\subsection{Skew-gradient projection}

The concept of skew-gradient projection $\xi _{s}(\xi )$ of canonical variables 
$\xi$ onto constraint submanifold plays very
important role in the Moyal quantization of constraint systems.
Geometrically, skew-gradient projection acts along phase flows 
$Id \mathcal{G}^{a}(\xi)$ generated by constraint functions. 
These flows are commutative in virtue of Eqs.(\ref{SB}): Using 
Eqs.(\ref{SB}) and the Jacobi identity, one gets 
$\{\mathcal{G}^{a},\{\mathcal{G}^{b},f\}\} = \{\mathcal{G}^{b},\{\mathcal{G}^{a},f\}\}$ 
for any function $f$, so the point of intersection with
$\Gamma^{*}$ is unique. Skew-gradient projections are investigated 
in Refs. \cite{NAKA84} and independently in Refs. \cite{KRFA,KRFF}.

\vspace{7mm} 
\begin{figure}[!htb]
\begin{center}
\includegraphics[angle=0,width=5.0 cm]{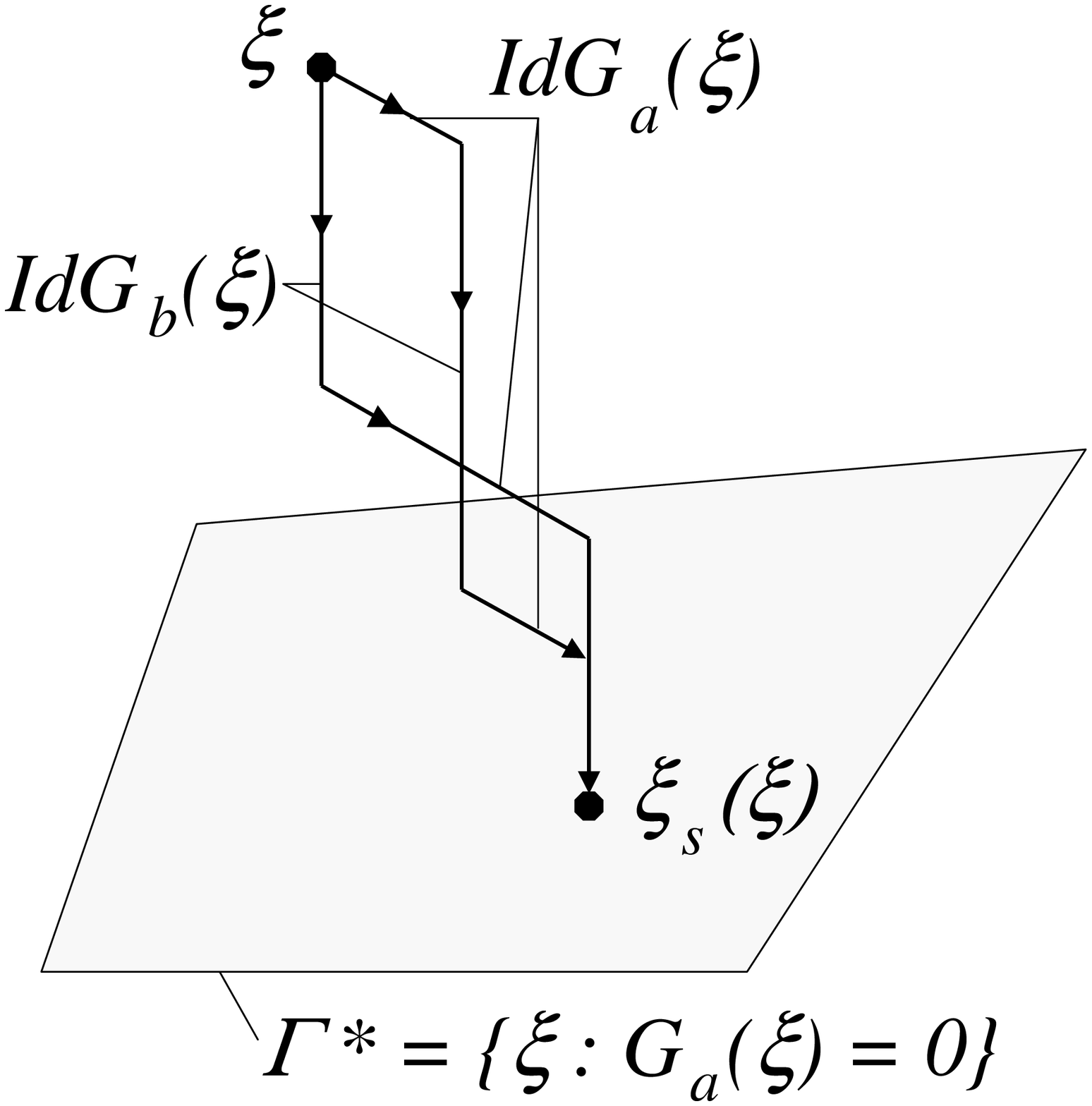}
\end{center}
\caption{Schematic presentation of skew-gradient projection onto constraint
submanifold along commuting phase flows generated by constraint functions.}
\label{fig1}
\end{figure}

To construct skew-gradient projections, we start from equations 
\begin{equation}
\{\xi _{s}(\xi ),\mathcal{G}_{a}(\xi )\}=0  \label{CG}
\end{equation}
which say that point $\xi _{s}(\xi ) \in \Gamma^{*}$ is left
invariant by phase flows generated by $\mathcal{G}_{a}(\xi )$. Using
symplectic basis (\ref{SB}) for the constraints and expanding 
\begin{equation}
\xi _{s}(\xi )=\xi +X^{a}\mathcal{G}_{a}+\frac{1}{2}X^{ab}\mathcal{G}_{a}%
\mathcal{G}_{b}+...
\end{equation}
in the power series of $\mathcal{G}_{a}$, one gets 
\begin{eqnarray}
\xi _{s}(\xi )=\sum_{k=0}^{\infty }\frac{1}{k!}\{...\{\{\xi ,\mathcal{G}%
^{a_{1}}\},\mathcal{G}^{a_{2}}\},...\mathcal{G}^{a_{k}}\}  \nonumber \\
\times \mathcal{G}_{a_{1}}\mathcal{G}_{a_{2}}...\mathcal{G}_{a_{k}}.
\label{SGRAD}
\end{eqnarray}
Similar projection can be made for function $f(\xi )$: 
\begin{eqnarray}
f_{s}(\xi )=\sum_{k=0}^{\infty }\frac{1}{k!}\{...\{\{f(\xi ),\mathcal{G}%
^{a_{1}}\},\mathcal{G}^{a_{2}}\},...\mathcal{G}^{a_{k}}\}  \nonumber \\
\times \mathcal{G}_{a_{1}}\mathcal{G}_{a_{2}}...\mathcal{G}_{a_{k}}.
\label{FSG}
\end{eqnarray}
It satisfies
\begin{equation}
f_{s}(\xi )=f(\xi _{s}(\xi )).  \label{FSSF}
\end{equation}
Constraint functions are in involution with projected function:
\begin{equation}
\{ f_{s}(\xi ), \mathcal{G}_{a}(\xi) \}=0.  \label{involution}
\end{equation}
Consequently, $f_{s}(\xi )$ does not vary along $Id\mathcal{G}_{a}(\xi)$, since
\[
\{ f(\xi ), g(\xi) \} \equiv \frac{\partial f(\xi)}{\partial \xi^{i}} (Idg(\xi))^{i}.
\]

Applying Eqs.(\ref{FSG}) and (\ref{FSSF}) to constraint functions $\mathcal{G%
}_{a}(\xi)$, one concludes that the point $\xi _{s}(\xi )$ belongs to the
constraint submanifold 
\begin{equation}
\mathcal{G}_{a}(\xi _{s}(\xi ))=0.  \label{on}
\end{equation}
The constraint submanifold can therefore be described equivalently
as $\Gamma^{*} = \{\xi _{s}(\xi ) : \xi \in T_{*}\mathbb{R}^{n}\}$.

An average of function $f(\xi)$ is calculated using the probability
density distribution $\rho(\xi)$ and the Liouville measure restricted to the
constraint submanifold \cite{FADD}: 
\begin{equation}
<f>=\int \frac{d^{2n}\xi}{(2\pi)^{n}} (2\pi)^{m}\prod_{a=1}^{2m} \delta (%
\mathcal{G}_{a}(\xi)) f(\xi)\rho(\xi).  \label{FAMEASURE}
\end{equation}
On the constraint submanifold $\xi _{s}(\xi ) = \xi$, so $f(\xi )$ and $%
\rho(\xi)$ can be replaced with $f_{s}(\xi )$ and $\rho_{s}(\xi )$.

There exist therefore equivalence classes of functions in phase space: 
\begin{equation}
f(\xi )\sim g(\xi )\leftrightarrow f_{s}(\xi )=g_{s}(\xi ).  \label{EQ}
\end{equation}
The symbol $\sim$ means that functions are equal in the weak
sense, $f(\xi )\approx g(\xi )$, i.e., on the constraint submanifold. We
shall see that symbols $\sim $ and $\approx $ acquire distinct meaning
upon quantization. Note that $f(\xi )\sim f_{s}(\xi ).$ Eqs.(\ref{FSSF}) and
(\ref{on}) imply $\mathcal{G}_{a}\sim 0$. Constraint functions belong to an
equivalence class containing zero.

\subsection{Dirac bracket in terms of Poisson bracket on constraint submanifold}

Given hamiltonian function $\mathcal{H}$, the evolution of function $f$
is described using the Dirac bracket \cite{DIRAC}
\begin{equation}
\frac{\partial }{\partial t}f=\{f,{}\mathcal{H}\}_{D}.  \label{EV}
\end{equation}
In the symplectic basis (\ref{SB}), the Dirac bracket looks like 
\begin{equation}
\{f,g\}_{D}=\{f,g\}+\{f,\mathcal{G}^{a}\}\{\mathcal{G}_{a},g\}.  \label{DBNA}
\end{equation}
On the constraint submanifold, one has 
\begin{equation}
\{f,g\}_{D}=\{f,g_{s}\}=\{f_{s},g\}=\{f_{s},g_{s}\}.  \label{DB1}
\end{equation}
Calculation of the Dirac bracket can be replaced therefore with
calculation of the Poisson bracket for functions projected onto the
constraint submanifold.

Two functions are equivalent provided they coincide on the constraint
submanifold. The hamiltonian functions determine the evolution of systems
and play thereby special role. Two hamiltonian functions are equivalent if
they generate within $\Gamma^{*}$ phase flows whose
projections onto the tangent plane of the constraint submanifold are
identical. One may suppose that the equivalence relation for functions,
defined above, does not apply to hamiltonian functions, since skew-gradients
of hamiltonian functions enter the problem either. This is not the case,
however. The components of the hamiltonian phase flow, which belong to a
subspace spanned at $\Gamma^{*}$ by phase flows of the
constraint functions, do not affect dynamics and could be different,
whereas the skew-gradient projection (\ref{FSG}) \textit{does not modify
components of skew-gradients of functions, tangent to constraint
submanifold}. We illustrate it schematically on Fig. 2. The geometrical
sense of the Dirac bracket reduces to dropping the component of the hamiltonian
phase flow which does not belong to tangent plane of the constraint
submanifold. Equivalently, those components can be made to vanish with the
help of the skew-gradient projection. $\mathcal{H}$ and $\mathcal{H}_{s}$
are thereby \textit{dynamically equivalent}, so Eq.(\ref{EQ}) characterizes
an equivalence class for the hamiltonian functions either. Among functions
of this class, $\mathcal{H}_{s}$ is the one whose phase flow is
skew-orthogonal to phase flows of the constraint functions, i.e., 
$\{\mathcal{G}_{a},\mathcal{H}_{s}\} = (Id\mathcal{G}_{a},Id\mathcal{H}_{s}) = 0$.

\vspace{6mm} 
\begin{figure}[!htb]
\begin{center}
\includegraphics[angle=0,width=5.0 cm]{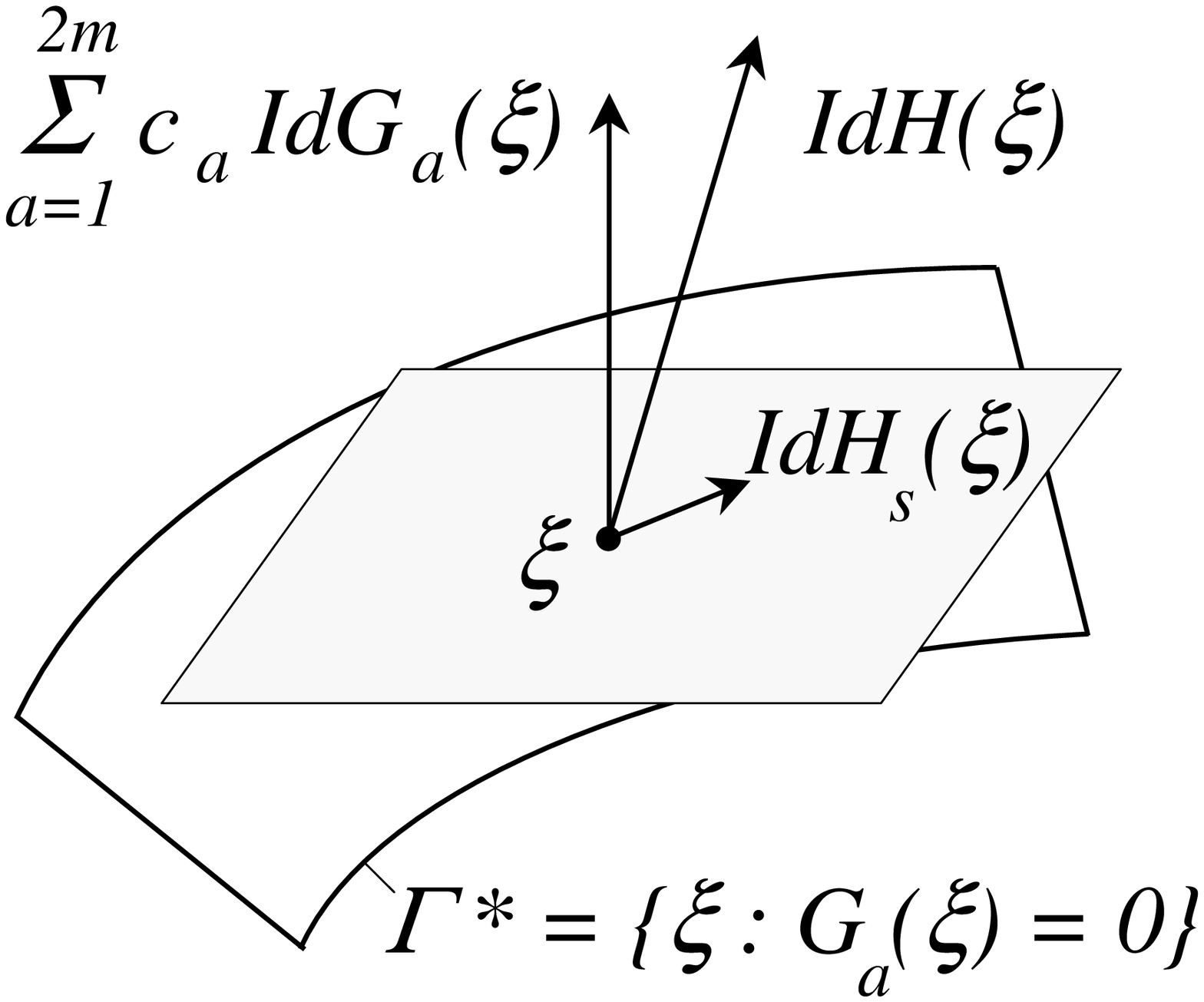}
\end{center}
\caption{Schematic presentation of phase flows $Id\mathcal{H}(\xi)$ and $Id%
\mathcal{H}_{s}(\xi)$ generated by hamiltonian function $\mathcal{H}(\xi)$
and projected hamiltonian function $\mathcal{H}_{s}(\xi)$ at point $\xi$
of constraint submanifold $\Gamma^{*}$. 
The phase flow $Id\mathcal{H}_{s}(\xi)$ belongs to the tangent plane of 
$\Gamma^{*}$. The hamiltonian phase flow $Id\mathcal{H}(\xi)$ admits decomposition
$Id\mathcal{H}(\xi) = \sum_{a=1}^{2m}c_{a}Id\mathcal{G}^{a}(\xi) + 
Id\mathcal{H}_{s}(\xi)$. Within the constraint submanifold (i.e. $\xi \in \Gamma^{*}$ 
and $\xi + d\xi \in \Gamma^{*}$) one has 
$d\mathcal{G}^{a}(\xi) = 0$ and therefore $0 = d\xi^{i}
\partial \mathcal{G}^{a}(\xi)/ \partial \xi^{i}  = (Id\mathcal{G}^{a}(\xi),d\xi)$.
The first term $\sum_{a=1}^{2m}c_{a}Id\mathcal{G}^{a}(\xi)$ is therefore skew-orthogonal
to any vector $d\xi$ of the tangent plane. 
}
\label{fig2}
\end{figure}

Replacing $\mathcal{H}$ with $\mathcal{H}_{s}$, one can rewrite the
evolution equation in terms of the Poisson bracket (cf. Eq.(\ref{EV})): 
\begin{equation}
\frac{\partial }{\partial t}f=\{f,{}\mathcal{H}_{s}\}.  \label{PEV}
\end{equation}
The evolution does not mix up the equivalence classes. 

\textit{The physical observables in second-class constraints systems are
associated with the equivalence classes of real functions in the
unconstrained phase space. The equivalence classes constitute a vector space 
}$\mathcal{O}$\textit{\ equipped with two multiplication operations, the
associative pointwise product and the skew-symmetric Dirac bracket $%
\{,\}_{D},$ which confer }$\mathcal{O}$\textit{\ a Poisson algebra structure.%
}

Instead of working with equivalence classes of functions $\mathcal{E}_{f}$,
one can work with their representatives $f_{s}$ defined uniquely by the
skew-gradient projection. The one-to-one mapping $\mathcal{E}%
_{f}\leftrightarrow f_{s}$ induces a Poisson algebra structure on the set of
projected functions. The sum $\mathcal{E}_{f}+\mathcal{E}_{g}$ converts to $%
f_{s}+g_{s}$, the associative product $\mathcal{E}_{f}\mathcal{E}_{g}$
converts to the pointwise product $f_{s}g_{s}$, while the Dirac bracket
becomes the Poisson bracket: 
\begin{equation}
\{f_{s},g_{s}\}_{D}=\{f_{s},g_{s}\}.  \label{DBasPB}
\end{equation}
These operations satisfy the Leibniz' law and the Jacobi identity and, since 
$(f_{s}+g_{s})_{s}=f_{s}+g_{s}$, $(f_{s}g_{s})_{s}=f_{s}g_{s}$, and $%
\{f_{s},h_{s}\}_{s}=\{f_{s},h_{s}\}$, keep the set of projected functions
closed.

\subsection{Dirac bracket in terms of Poisson bracket on and outside of 
constraint submanifold}

Outside of the constraint submanifold functions do not make any physical
sense. It is sufficient thus to work with the Dirac bracket on the
constraint submanifold. The evolution problem in such a case can
consistently be formulated in terms of the Poisson bracket for functions
projected onto the constraint submanifold.

The Dirac bracket is, however, well defined in the whole phase space. Redefinition
of constraint functions by shifts 
$\mathcal{G}_{a}(\xi ) \rightarrow \mathcal{G}_{a}(\xi ) + \mathrm{constant}$ 
leaves the Dirac bracket unchanged, because it depends on derivatives of constraint
functions only. It is not the case for the Poisson bracket applied to
projected functions. This is why Eq.(\ref{DB1}) is valid on constraint
submanifold only.

One can modify projection formalism to fit the above-mentioned property
of the Dirac bracket. Suppose we wish to find the Dirac bracket of functions 
$f(\zeta )$ and $g(\zeta )$ at a point $\zeta =\xi $ outside of the
constraint submanifold. The intersection of level sets $\{ \zeta : \mathcal{G}%
_{a}(\zeta )=\mathcal{G}_{a}(\xi ) \}$ can be
considered as new constraint submanifold defined by constraint functions
\[
\Delta \mathcal{G}_{a}(\zeta )=\mathcal{G}_{a}(\zeta )-\mathcal{G}_{a}(\xi
).  
\]
Projected functions depend thereby on both $\zeta $ and $\xi $: 
\begin{eqnarray}
f_{S}(\zeta ) &=& \sum_{k=0}^{\infty }\frac{1}{k!}\{...\{\{f(\zeta ),\Delta 
\mathcal{G}^{a_{1}}\},\Delta \mathcal{G}^{a_{2}}\},...\Delta \mathcal{G}%
^{a_{k}}\}  \nonumber \\
&\times& \Delta \mathcal{G}_{a_{1}}\Delta \mathcal{G}_{a_{2}}...\Delta 
\mathcal{G}_{a_{k}}  \label{PROOUT}
\end{eqnarray}
and similarly for $g(\zeta ).$ The Poisson brackets
are calculated with respect to $\zeta $ while $\xi$ is a parameter. 
The appropriate extension looks like 
\begin{eqnarray}
\{f(\xi ),g(\xi )\}_{D} &=& \{f(\zeta),g_{S}(\zeta )\}|_{\zeta = \xi}  \nonumber \\
&=& \{f_{S}(\zeta ),g(\zeta )\}|_{\zeta = \xi}  \nonumber \\
&=& \{f_{S}(\zeta ),g_{S}(\zeta )\}|_{\zeta = \xi}.
\label{OUTSIDE}
\end{eqnarray}
In Eq.(\ref{DB1}) all four terms are pairwise distinct functions in the whole 
phase space. These functions coincide on the constraint submanifold only. 
In Eq.(\ref{OUTSIDE}) all four terms coincide in the whole phase space.
If $\xi \in \Gamma^{*}$, we reproduce the result (\ref{DB1}) derived earlier. 


\section{Quantum constraint systems in phase space}
\setcounter{equation}{0} 


Scheme presented in the previous Sect. 3 is suitable to approach description
of quantum constraint systems in phase space. We give final results and refer
to \cite{KRFA} for intermediate steps.

We remind that classical hamiltonian function $\mathcal{H}(\xi)$ and constraint 
functions $\mathcal{G}_{a}(\xi)$ are distinct in general from their quantum
analogues  ${H}(\xi)$ and ${G}_{a}(\xi)$. These dissimilarities are connected to
the usual ambiguities in quantization of classical systems, being not specific 
for the problem we are interested in. It is required only 
\begin{eqnarray}
\lim_{\hbar \rightarrow 0}{H}(\xi) &=& \mathcal{H}(\xi), \nonumber \\
\lim_{\hbar \rightarrow 0}{G}_{a}(\xi) &=& \mathcal{G}_{a}(\xi). \nonumber 
\end{eqnarray}
In what follows $\Gamma^{*} = \{\xi: {G}_{a}(\xi) = 0 \}$.

\subsection{Quantum deformation of the Dirac bracket on constraint submanifold}

The quantum constraint functions ${G}_{a}(\xi)$ satisfy 
\begin{equation}
{G}_{a}(\xi )\wedge {G}_{b}(\xi )=\mathcal{I}_{ab}.  \label{SBAS}
\end{equation}
In classical limit, ${G}_{a}(\xi)$ turn to $\mathcal{G}_{a}(\xi)$.

The quantum-mechanical version of the skew-gradient projections is defined
with the use of the Moyal bracket 
\begin{equation}
\xi _{t}(\xi )\wedge {G}_{a}(\xi )=0.  \label{CG3}
\end{equation}

The projected canonical variables have the form 
\begin{eqnarray}
\xi _{t}(\xi )&=&\sum_{k=0}^{\infty }\frac{1}{k!}(...((\xi \wedge {G}%
^{a_{1}}) \wedge {G}^{a_{2}})...\wedge {G}^{a_{k}})  \nonumber \\
&&\circ {G}_{a_{1}}\circ {G}_{a_{2}}...\circ {G}_{a_{k}}.  \label{SGRAD3}
\end{eqnarray}
The quantum analogue of Eq.(\ref{FSG}) is 
\begin{eqnarray}
f_{t}(\xi )&=&\sum_{k=0}^{\infty }\frac{1}{k!}(...((f(\xi )\wedge {G}%
^{a_{1}})\wedge {G}^{a_{2}})...\wedge {G}^{a_{k}})  \nonumber \\
&&\circ {G}_{a_{1}}\circ {G}_{a_{2}}...\circ {G}_{a_{k}}.  \label{SGRAD4}
\end{eqnarray}
The function $f_{t}(\xi)$ obeys equation
\begin{equation}
f_{t}(\xi ) \wedge {G}_{a}(\xi) = 0.
\label{LAB}
\end{equation}

The evolution equation which is the analogue of Eq.(\ref{PEV}) takes the
form 
\begin{equation}
\frac{\partial}{\partial t}f(\xi) = f(\xi) \wedge {H}_{t}(\xi)  \label{PEV2}
\end{equation}
where ${H}_{t}(\xi)$ is the hamiltonian function projected onto the
constraint submanifold as prescribed by Eq.(\ref{SGRAD4}). Taking projection 
of Eq.(\ref{PEV2}) we get evolution equation in the closed form for projected 
functions:
\begin{equation}
\frac{\partial}{\partial t}f_{t}(\xi) = f_{t}(\xi) \wedge {H}_{t}(\xi)  \label{PEV7}
\end{equation}

\textit{The quantum deformation of the Dirac bracket represents the Moyal
bracket for two functions projected quantum-mechanically onto the constraint
submanifold.}

The formal structure of the dynamical quantum system is described by the
scheme (\ref{SCHEME}) with the word "functions" replaced by the phrase "projected
functions" and $f$ and $g$ replaced by $f_t$ and $g_t$, respectively. The
star-product is an associative operation, whereas the Moyal bracket for
projected functions satisfies the Leibniz' law and, respectively, the Jacobi
identity.

Projected functions in phase space are objects associated to quantum
observables. Functions which have the same projections are physically
equivalent. We can unify such functions into equivalence classes. The
star-product and the Moyal bracket for projected functions generate 
for equivalence classes a Poisson algebra structure accordingly.

\begin{table}[tbp]
\caption{ Brackets which govern evolution in phase space of
functions (second column) and projected functions (third column) of
classical systems (first row) and quantum systems (second row). The right
upper corner shows the Dirac bracket expressed in terms of the Poisson
bracket of functions projected onto the constraint submanifold. The left
upper corner is the Poisson bracket. The left lower corner is the Moyal
bracket, which represents the quantum deformation of the Poisson bracket.
The operation $f_{t}\wedge g_{t} $ is the quantum deformation of the Dirac
bracket. }
\label{lab2}
\begin{center}
\begin{tabular}{|l|c|c|}
\hline
Systems: & unconstrained & constrained \\ \hline
classical & $\{f,g\} $ & $\{f_{s},g_{s}\}$ \\ \hline
quantum & $f\wedge g$ & $f_{t}\wedge g_{t} $ \\ \hline
\end{tabular}
\end{center}
\par
\vspace{-2mm}
\end{table}

The bracket $f_{t} \wedge g_{t}$ constructed in \cite{KRFA} gives the
deformation of the Dirac bracket on $\Gamma^{*}$. What about the whole phase space?

\subsection{Quantum deformation of the Dirac bracket on and outside of constraint submanifold}

One can generalize the operation $f_{t}\wedge g_{t}$ to match in classical
limit the Dirac bracket outside of the constraint submanifold. We can
proceed like in the classical case by writing projected functions in the
form 
\begin{eqnarray}
f_{T}(\zeta ) &=& \sum_{k=0}^{\infty }\frac{1}{k!}(...( (f(\zeta ) \wedge
\Delta {G}^{a_{1}}) \wedge \Delta {G}^{a_{2}}) \wedge  \nonumber \\
&...& \Delta {G}^{a_{k}}) \circ \Delta {G}_{a_{1}} \circ \Delta {G}%
_{a_{2}}... \circ \Delta {G}_{a_{k}}  \label{PROOUTQUA}
\end{eqnarray}
where 
\[
\Delta {G}_{a}(\zeta ) = {G}_{a}(\zeta ) - {G}_{a}(\xi ). 
\]
The Moyal brackets and the $\circ $-products entering this equation are 
calculated with respect to $\zeta $. The desired extension looks like 
\begin{equation}
f_{T}(\zeta ) \wedge g_{T}(\zeta )|_{\zeta = \xi}.
\label{OUTSIDEQUA}
\end{equation}
It is assumed that the constraint functions ${G}_{a}(\xi)$ satisfy the bracket
relations (\ref{SBAS}) at $\xi \notin \Gamma^{*}$. Expression (\ref{OUTSIDEQUA}) is valid on
and outside of the constraint submanifold. If $\xi \in \Gamma^{*}$, we reproduce 
operation $f_{t}(\xi) \wedge g_{t}(\xi)$ announced earlier.

\subsection{Completeness of the set of projected operators of canonical 
coordinates and momenta}

The set of operators $\mathfrak{x}^{i}$ is known to be complete, so that any 
operator $\mathfrak{f}$ can be represented as a symmetrized (probably 
infinite) weighted sum of products of operators $\mathfrak{x}^{i}$. 
In the sense of the Taylor expansion, one can write $\mathfrak{f} = f(\mathfrak{x})$. 
The one-to-one correspondence between operators $\mathfrak{f} \in Op(L^{2}(\mathbb{R}^{n}))$ and 
functions in phase space $f(\xi)$, based on the Taylor expansion, is equivalent to the Weyl's 
association rule.

The similar completeness condition holds for projected operators of canonical 
variables $\mathfrak{x}^{i}_{t}$ which are inverse Weyl's transforms of $\xi_{t}^{i}(\xi)$. 
Apparently, any operator $\mathfrak{f}$ acting in the Hilbert space can be represented 
as an operator function $\varphi (\mathfrak{G}^{a},\mathfrak{x}^{i}_{t})$. Applying 
projection to the symmetrized product of $k$ constraint operators $\mathfrak{G}^{a}$, 
which are inverse Weyl's transforms of $G^{a}(\xi)$, one gets a series 
like $1 - k + \frac{1}{2!}k(k-1) +\ldots = (1 - 1)^{k} = 0$, and so
\begin{equation}
(\mathfrak{G}^{(a_{1}}\mathfrak{G}^{a_{2}}...\mathfrak{G}^{a_{k})})_{t} = 0.
\label{GGGGT}
\end{equation}
The Taylor series of $\varphi (\mathfrak{G}^{a},\mathfrak{x}^{i}_{t})$ generates thereby 
vanishing terms involving $\mathfrak{G}^{a}$. We thus obtain 
\begin{equation}
(\varphi(\mathfrak{G}^{a},\mathfrak{x}^{i}_{t}))_{t} = \varphi(0,\mathfrak{x}^{i}_{t}).
\label{GGGGTGREAT}
\end{equation}
Respectively, any function projected quantum-mechanically onto the constraint submanifold 
can be represented in the form
\begin{equation}
f_{t}(\xi) = \varphi(\star \xi_{t}(\xi)).
\label{GREAT}
\end{equation}
One can pass to classical limit to get Eq.(\ref{FSSF}). Constructing $\varphi(\xi)$
from $f(\xi)$ is a non-trivial task equivalent to solving constraints. The operator counterpart 
of Eq.(\ref{GREAT}),
\begin{equation}
\mathfrak{f}_{t} = \varphi(\mathfrak{x}_{t}),
\label{3947}
\end{equation}
demonstrates the completeness of projected set of operators
of canonical coordinates and momenta. Accordingly, Eq.(\ref{GREAT}) shows 
completeness of the set of $\xi_{t}^{i}(\xi)$ in description of projected 
functions. It is worthwhile to notice that Eq.(\ref{GGGGT}) does not extend to 
antisymmetric products of $\mathfrak{G}^{a}$ as one sees from 
$[\mathfrak{G}^{a},\mathfrak{G}^{b}]_{t} = (- \mathcal{I}^{ab})_{t} 
= - \mathcal{I}^{ab} \neq [\mathfrak{G}^{a}_{t},\mathfrak{G}^{b}_{t}] = 0$ 
where condition $\mathfrak{G}^{a}_{t} = 0$ is taken into account.

\section{Conclusion}
\setcounter{equation}{0} 

We made short introduction to the Weyl's association rule and the Groenewold star-product 
technique for unconstrained and constraint systems. The attention was focused to the
evolution problem.

A generalization of the quantum deformation of the Dirac bracket is
constructed to match smoothly classical Dirac bracket in the whole phase
space at $\hbar \rightarrow 0$.

The use of skew-gradient projection formalism allows to treat unconstrained and constraint 
systems essentially on the same footing. Projections of solutions of quantum evolution 
equations onto the constraint submanifold comprise the entire information on quantum 
dynamics of constraint systems. 


\end{document}